\documentclass[reprint,aps,prl,superscriptaddress,noeprint]{revtex4-2}

\usepackage[normalem]{ulem}
\usepackage{amsmath}
\usepackage{amsfonts}
\usepackage{amssymb} 
\usepackage{graphicx}
\usepackage{float}
\usepackage[caption=false]{subfig}
\usepackage{hyperref}
\hypersetup{
	colorlinks=true,       
	linkcolor=blue,          
	citecolor=green,       
	filecolor=magenta,      
	urlcolor=blue          
}
\usepackage{siunitx}
\usepackage{booktabs}
\usepackage{multirow}
\usepackage{dblfloatfix}

\usepackage[dvipsnames]{xcolor}

\begin{document}
\title{Simultaneous nondestructive measurement of many polar molecules using Rydberg atoms}	

\author{Jeremy T. Young}
\email[Corresponding author: ]{j.t.young@uva.nl}
\affiliation{Institute of Physics, University of Amsterdam, 1098 XH Amsterdam, the Netherlands}

\author{Kang-Kuen Ni}
\affiliation{Department of Chemistry and Chemical Biology, Harvard University, Cambridge, Massachusetts 02138, USA}
\affiliation{Department of Physics, Harvard University, Cambridge, Massachusetts 02138, USA}
\affiliation{Harvard-MIT Center for Ultracold Atoms, Cambridge, Massachusetts 02138, USA}

\author{Alexey V. Gorshkov}
\affiliation{Joint Quantum Institute and Joint Center for Quantum Information and Computer Science, NIST/University of Maryland, College Park, Maryland 20742 USA}

\date{\today}

\begin{abstract}
Tweezer arrays of polar molecules present new opportunities for quantum science and quantum information. However, a major challenge, especially in bialkali molecule platforms, is the fact that current measurement schemes for the internal states are destructive. In this work, we present a method to use Rydberg atoms to nondestructively measure the internal state of a molecular qubit. We achieve this via microwave dressing of both molecules and Rydberg atoms, allowing us to tune the interactions so that there are minimal Rydberg-Rydberg interactions and many measurements can take place simultaneously. We consider two experimentally-motivated examples of detecting $^{23}$Na$^{133}$Cs and $^{87}$Rb$^{133}$Cs with $^{133}$Cs atoms. Finally, we discuss several strategies for mitigating various sources of crosstalk.

\end{abstract}


\maketitle

In the past decade, tweezer arrays of neutral atoms have rapidly positioned themselves as one of the leading quantum platforms due to their configurability and high fidelity gates \cite{Endres2016,Barredo2016,Browaeys2020,Kaufman2021,Bluvstein2024,Manetsch2024}. Parallel to these advances, this approach has been applied to polar molecules \cite{Liu2018,Anderegg2019,He2020a,Zhang2022, Bao2023,Holland2023b,Guttridge2023,Vilas2024, Picard2024a, Ruttley2025}, opening the door to new possibilities for quantum computing and simulation \cite{Ni2018,Cornish2024}. However, an important hurdle for tweezer arrays of bialkali polar molecules is state detection. In contrast to alkali atoms, alkaline-earth atoms, and $^2 \Sigma$ molecules, bialkali molecules do not possess a cycling transition that can be used for nondestructive measurements of the internal molecular states. As a consequence, current experiments measure the internal states via a destructive procedure where the molecule is converted to a pair of atoms depending on the state of the molecule, both in tweezer arrays \cite{Ruttley2024,Picard2024} and optical lattices \cite{Covey2018,Christakis2023,Rosenberg2022,Tobias2022}. In order to perform additional operations, additional molecules must be utilized, limiting the scalability of the platform. 

\begin{figure}[h!]
    \centering
    \includegraphics[width=1\linewidth]{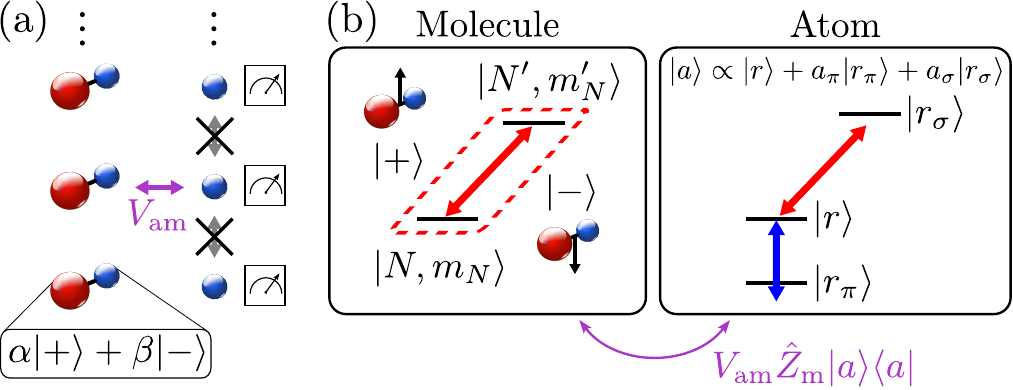}
    \caption{(a) Each molecule (left) is paired with a nearby atom (right) with interaction $V_{\text{am}}$, entangling each pair. The internal molecular state $\alpha |+\rangle + \beta |-\rangle$ is then measured by measuring the internal state of the entangled atom. (b) Example of dressing scheme. The molecules are dressed resonantly with circular polarization, producing eigenstates $|\pm\rangle \equiv (|N,m_N\rangle \pm |N',m_N'\rangle)/\sqrt{2}$, where $N,m_N$ are the quantum numbers for the molecules rotational angular momentum and its $z$-projection and $m_N' = m_N+1$ in this example. The molecular drive also couples the Rydberg states $|r\rangle, |r_\sigma\rangle$ of the atom, which is subject to an additional drive with linear polarization that couples $|r\rangle$ and $|r_\pi\rangle$. With proper choice of the drives, the $|a\rangle$ eigenstate does not interact with itself but interacts with the molecule. }
    \label{fig:Fig1}
\end{figure}

Various approaches have been proposed for nondestructive measurements of bialkali polar molecules, including dispersive measurements using anisotropic polarizability of the molecule \cite{Guan2020} and through coupling to Rydberg atoms \cite{Kuznetsova2016,Zeppenfeld2017,Wang2022,Zhang2022a}. One complication for the use of Rydberg atoms as a means of measuring the molecular states is the strong Rydberg-Rydberg interaction, which prevents the simultaneous measurement of multiple nearby molecules. Here, we propose to overcome this difficulty by utilizing microwave-dressed asymmetric blockade \cite{Young2021}. This is achieved by using multiple microwave fields to engineer Rydberg interactions which destructively interfere with one another, resulting in dressed Rydberg states which do not interact with one another. Through a proper tuning of the dressing parameters, the molecule-Rydberg interaction can likewise be simultaneously engineered such that the internal states of a molecule-Rydberg pair become entangled, enabling a measurement of the molecular state by measuring the atom's internal state, see Fig.~\ref{fig:Fig1}. Furthermore, local measurement capabilities are determined by the local control of the atoms rather than the molecules.

\emph{Nondestructive measurement.}---First, we discuss the general scheme for realizing the nondestructive measurement, which relies on entangling the internal states of the atom and the molecule.
We start with the case of a single molecule-atom pair, where we may ignore any crosstalk with other atoms or molecules. We consider a bialkali molecule in the electro-vibrational ground state, which has Hamiltonian $\hat H_0 = B_0 \hat{\mathbf{N}}^2$, where $B_0$ is the rotational constant and $\hat{\mathbf{N}}$ the angular momentum operator of the molecule's rotations. We encode the qubit via a pair of rotational states $|N,m_N\rangle, |N',m_N'\rangle$ with $N' = N \pm 1$, defined by $\hat{\mathbf{N}}^2$ ($N$) and its $z$-projection $\hat{\mathbf{N}}_z$ ($m_N$). Specifically, we use equal superpositions of these two states
\begin{equation}
    |\pm\rangle = \frac{|N,m_N\rangle \pm |N',m_N'\rangle}{\sqrt{2}}.
\end{equation}
In practice, the qubits will be stored in a separate set of non-dressed states before the measurement (e.g.~hyperfine states), which are then mapped to $|\pm\rangle$ for the measurement. When the two rotational states have a nonzero transition dipole moment $\mu_{\text{m}}$ between one another, the $|\pm\rangle$ states have effective permanent dipole moments $\pm \mu_{\text{m}} \neq 0$ (and no transition dipole moment between one another). 

To detect the state of the molecule, we utilize the dipolar interaction between a Rydberg state $|a\rangle$ with (effective) permanent dipole moment $\mu_{\text{a}}$. The dipolar molecule-Rydberg interaction takes the form
\begin{equation}
    \hat V_{\text{am}} = c_{\wp} \mu_{\text{m}} \mu_{\text{a}} \frac{1-3\cos^2 \theta_{\text{am}}}{r_{\text{am}}^3} \hat Z_\text{m} |a \rangle \langle a|,
    \label{eq:interaction}
\end{equation}
where $r_{\text{am}}$ is the distance between the atom and molecule, $\theta_{\text{am}}$ is the angle between the two relative to the quantization axis, which we choose to be the $z$-axis, and $\hat Z_\text{m}$ is the Pauli-Z operator in the molecular $|\pm\rangle$ basis which arises due to the opposite sign of the permanent dipole moments. The factor $c_{\wp}$ is determined by the polarization ${\wp}=\pi,\sigma$ of the driven transition ($c_\pi =1, c_\sigma = -1/2$), which can be understood semi-classically as the time average of two dipoles in parallel ($\pi$) vs.~two dipoles rotating together in the $x$-$y$ plane ($\sigma$). For simplicity, we assume $\theta_{\text{am}} = \pi/2$.

\begin{figure}
    \centering
    \includegraphics[width=1\linewidth]{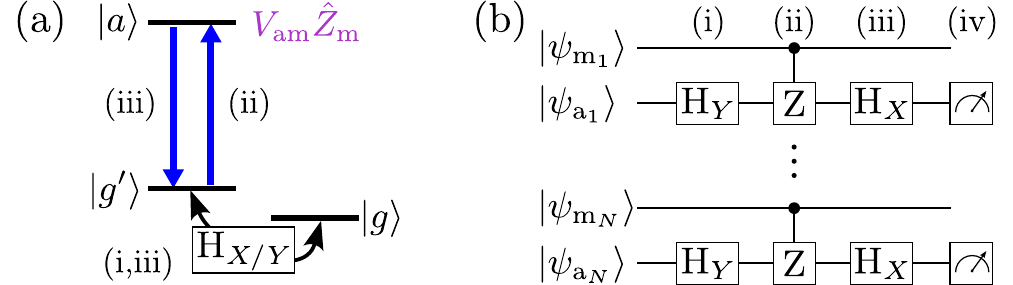}
    \caption{Schematic of nondestructive measurement scheme. (a) Atomic pulse sequence used to entangle atom-molecule pairs via atom-molecule interaction $V_{\text{am}} \hat Z_\text{m} |a \rangle \langle a|$ for Rydberg state $|a\rangle$ and long-lived states $|g\rangle, |g'\rangle$, where the atoms are initialized in $|g\rangle$. (b) Effective quantum circuit for molecular ($|\psi_{\text{m}_i} \rangle$) and atomic ($|\psi_{\text{a}_i}\rangle$) wavefunction pairs realized via pulse sequence which is composed of Hadamard gates $\textsc{H}_{X/Y}$, controlled-Z gates \textsc{CZ} enabled by the interaction, and measurements. See text for additional details on steps (i)-(iv).}
    \label{fig:schematic}
\end{figure}

Nondestructive measurement of a molecule in the state $\alpha |+\rangle + \beta |-\rangle$ can be realized as follows: (i) The atom is initialized in $|g\rangle$, upon which we apply a modified Hadamard gate which maps $|g\rangle \to (|g\rangle + i |g'\rangle )/\sqrt{2}$, where $|g\rangle,|g'\rangle$ are two long-lived states (e.g., ground states) of the atom. We denote this Hadamard gate as $H_{Y}$ since it maps to the Pauli-Y basis. (ii) The $|g'\rangle$ state is excited to $|a\rangle$, after which the atom and molecule interact for a time $ t = \left| \pi/[2 c_{\wp} \mu_{\text{m}} \mu_{\text{a}}] \right|$, which maps $i |\pm \rangle |a\rangle \to \pm |\pm \rangle |a\rangle$, realizing a controlled-Z gate. (iii) The $|a\rangle$ state is de-excited to $|g'\rangle$ and a conventional Hadamard gate (denoted $H_X$) is applied in $|g\rangle, |g'\rangle$, resulting in a final entangled state $\alpha |+\rangle |g\rangle + \beta |-\rangle |g' \rangle$. (iv) The atomic state is measured, thereby uniquely determining the state of the molecule. This process is depicted schematically in Fig.~\ref{fig:schematic}. Alternatively, spectroscopic measurement of the energy shift of the Rydberg state could provide another means for measuring the molecular state (see, e.g., \cite{Blatt2009, Zhu2025}).

The remainder of the text will focus on simultaneously preparing $|\pm\rangle, |a\rangle$ states which realize the interactions of Eq.~(\ref{eq:interaction}) via microwave dressing and minimizing many-body crosstalk. The strongest unwanted interactions are the Rydberg-Rydberg interactions. To reduce the role of these interactions, we use microwave dressing to engineer asymmetric interactions \cite{Young2021}. To reduce the role of the weakened Rydberg-Rydberg interactions as well as unwanted molecule-Rydberg and molecule-molecule interactions, we propose a variety of techniques, including a spin-echo-like approach which can, in principle, eliminate all unwanted interactions.

\emph{Asymmetric interactions.}---For the atom, we engineer a Rydberg state $|a\rangle$ which has dipolar interactions with the molecule but not other atoms, realizing a form of multi-species asymmetric interactions.
To do so, we couple three Rydberg states $|r\rangle,|r_\pi\rangle,|r_\sigma\rangle$ via microwave fields with Hamiltonian
\begin{equation}
\begin{aligned}
    \hat H_\text{mw,a} = &~ \Omega_\pi (|r \rangle \langle r_\pi | + |r_\pi \rangle \langle r|) - \Delta_\pi |r_\pi \rangle \langle r_\pi| \\
    & + \Omega_\sigma (|r \rangle \langle r_\sigma | + |r_\sigma \rangle \langle r|) - \Delta_\sigma |r_\sigma \rangle \langle r_\sigma|,
\end{aligned}
\end{equation}
where $\Omega_{\pi/\sigma}$ are the Rabi frequencies and $\Delta_{\pi/\sigma}$ the detunings for the drive on the $\pi$ or $\sigma$ transition. Furthermore, we require that one of the two microwave fields resonantly couples the rotational states of the molecule (which we denote with polarization ${\wp}= \pi,\sigma$), while the other field is off-resonant with any relevant molecular transitions 
\begin{equation}
\begin{aligned}
        \hspace{-.18cm} \hat H_{\text{mw,m}} &= \Omega_{\text{m}} (|N,m_N\rangle \langle N',m_N'| + |N',m_N'\rangle \langle N,m_N|) \\
        &= \Omega_{\text{m}} (|+\rangle \langle +| - |-\rangle \langle -|),
        \end{aligned}
\end{equation}
with an associated Rabi frequency $\Omega_{\text{m}}$ and eigenstates $|\pm\rangle$, thereby realizing the molecular qubit states as eigenstates of the microwave drive, see Fig.~\ref{fig:Fig1}(b) for ${\wp}=\sigma$. Depending on $\Omega_{\text{m}}$, hyperfine coupling from the nuclear spins can become important. In the case of stretched states (i.e., $m_N = N, m_N' = N'$ and analogous maximal $z$-projection values for the nuclear spins), which necessitate driving a ${\wp} =\sigma$ transition, the hyperfine coupling only shifts the energies of the molecular states, so we assume stretched states are used in the main text. In the End Matter, we discuss strategies for other nuclear states.

We consider resonant driving of the molecule for two reasons. First, it maximizes the effective permanent dipole moment of the molecular states. Second, it ensures that there are no molecule-atom interactions which can induce a transition between the two molecular dressed states, i.e., there is no transition dipole moment $|\pm\rangle \to |\mp\rangle$. 
Depending on which drive is resonant with the molecular transition, we fix the detuning for the corresponding Rydberg drive
\begin{equation}
    \Delta_{\wp} = \omega_{\text{m}} - \omega_{\wp},
\end{equation}
where $\omega_{\text{m}}$ is the transition frequency for the molecular states and $\omega_{\wp}$ is the transition frequency of the Rydberg states for the associated drive. In spite of this constraint, we can still ensure that the dipolar and van der Waals (vdW) Rydberg-Rydberg interactions are eliminated.

The Rydberg state can be expressed in the general form
\begin{equation}
    |a\rangle \propto |r\rangle + a_\pi |r_\pi\rangle +a_\sigma |r_\sigma\rangle,
\end{equation}
which interacts with other atoms in $|a\rangle$ with dipolar interaction strength \cite{Gorshkov2011,Young2021}
\begin{equation}
    V_{\text{aa,dd}} \propto |a_\pi|^2 \mu_\pi^2 - |a_\sigma|^2 \mu_\sigma^2/2,
    \label{eq:atomdipolar}
\end{equation}
with transition dipole moments $\mu_\pi = \langle r_\pi |d_0| r\rangle$ and $\mu_\sigma = \langle r_\sigma |d_{\pm 1}| r\rangle$ (the subscript of $d$ is determined by which transition is driven), where $d_p = \mathbf{\hat{e}_p} \cdot \mathbf{d} = d \sqrt{4 \pi/3} Y_p^1(\theta,\phi)$ ($p=0, \pm 1$) is a component of the dipole operator $\mathbf{d}$, $\mathbf{\hat{e}_0} = \mathbf{\hat{z}},\mathbf{\hat{e}_{\pm 1}} = \mp (\mathbf{\hat{x}}\pm i \mathbf{\hat{y}})/\sqrt{2}$, and $Y_p^1$ are spherical harmonics. Crucially, the sign difference of $\pi$ and $\sigma$ interactions means that with careful selection of the relative magnitudes of $a_\pi, a_\sigma$, this interaction can be tuned to 0 according to the constraint
\begin{equation}
    a_\sigma = \pm \sqrt{2} \mathcal{M} a_\pi, \quad \mathcal{M} \equiv |\mu_\pi/\mu_\sigma|,
\end{equation}
where we have assumed real-valued Rabi frequencies, and thus eigenvectors (see End Matter for a discussion of the effect of complex Rabi frequencies). Additionally, we assume that these phases are homogeneous due to the large wavelengths of the fields. 

In the rotating frame of the drives, the resultant atom-molecule dipolar interactions have strength $V_{\text{am,dd}} = C_{3,{\wp}}/r_{\text{am}}^3$ with dispersion coefficients
\begin{equation}
    C_{3,{\wp}} = \pm |c_{\wp}|^{1/2} \mu_{\text{m}} \mu_\pi \frac{a_\pi}{1+a_\pi^2(1+2 \mathcal{M}^2)},
\end{equation}
where $r_{\text{am}}$ is the atom-molecule separation, ${\wp}$ denotes the two options for the molecular drive polarization, and the sign of the interaction is opposite for the $|\pm\rangle$ states. Note that only the dipolar interactions associated with the nearly-resonant transitions of the molecule and atom contribute. The interactions in both cases are maximized with strength $C_{3,{\wp}} = \pm |c_{\wp}|^{1/2} \mu_{\text{m}} \mu_\pi/(2 \sqrt{1+2 \mathcal{M}^2})$ when $|a_\pi| = (1+2 \mathcal{M}^2)^{-1/2}$. We parameterize $a_\pi \equiv \alpha/\sqrt{1+2 \mathcal{M}^2}$ or via $\beta \equiv \alpha^{-1} - \alpha$, where the use of $\beta$ renders the interaction strength $C_{3,{\wp}} = \pm |c_{\wp}|^{1/2} \mu_{\text{m}} \mu_\pi /\sqrt{(1 + 2 \mathcal{M}^2)(4+\beta^2)}$ symmetric in $\beta \to - \beta$. Full expressions for the dressing in either case of molecular drive polarization are presented in the End Matter.

\emph{vdW Interactions}.---While the dipolar interactions between Rydberg atoms have been removed, there can still be strong interactions of the form 
\begin{equation}
\hat V_{\text{aa}}(r_{\text{aa}}) = \left(\frac{C_6}{r_{\text{aa}}^6} + \frac{C_9}{r_{\text{aa}}^9}+\frac{C_{12}}{r_{\text{aa}}^{12}}\right) |a a\rangle \langle a a|,    
\end{equation}
where $r_{\text{aa}}$ is the atom-atom distance, which we must account for. In addition to the vdW interactions ($C_6$), we include higher-order interactions (with dispersion coefficients $C_9, C_{12}$) from the perturbative expansion in anticipation of tuning $C_6$ to 0.

The dressed states can be parameterized via three free real parameters (two coefficients for each of the three states subject to three orthogonality conditions) and are left unchanged under rescaling of the dressing Hamiltonian, giving a fourth free parameter (corresponding to the four dressing parameters). However, since the detuning associated with the molecular drive is fixed and Eq.~(\ref{eq:atomdipolar}) must be satisfied, the two constraints leave two free parameters, which we take to be $\beta$ and the Rydberg Rabi frequency $\Omega_{{\wp}}$. By tuning $\beta, \Omega_{\wp}$, the vdW interactions can be removed through a similar process to the dipolar interactions of destructively interfering different contributions with different signs. In this case, tuning the two free parameters of the dressing corresponds to tuning the relevant transition dipole moments (due to the dressed basis) as well as the Rydberg state energies (due to the light shifts).

\begin{figure}[b]
    \centering
    \includegraphics[width=1\linewidth]{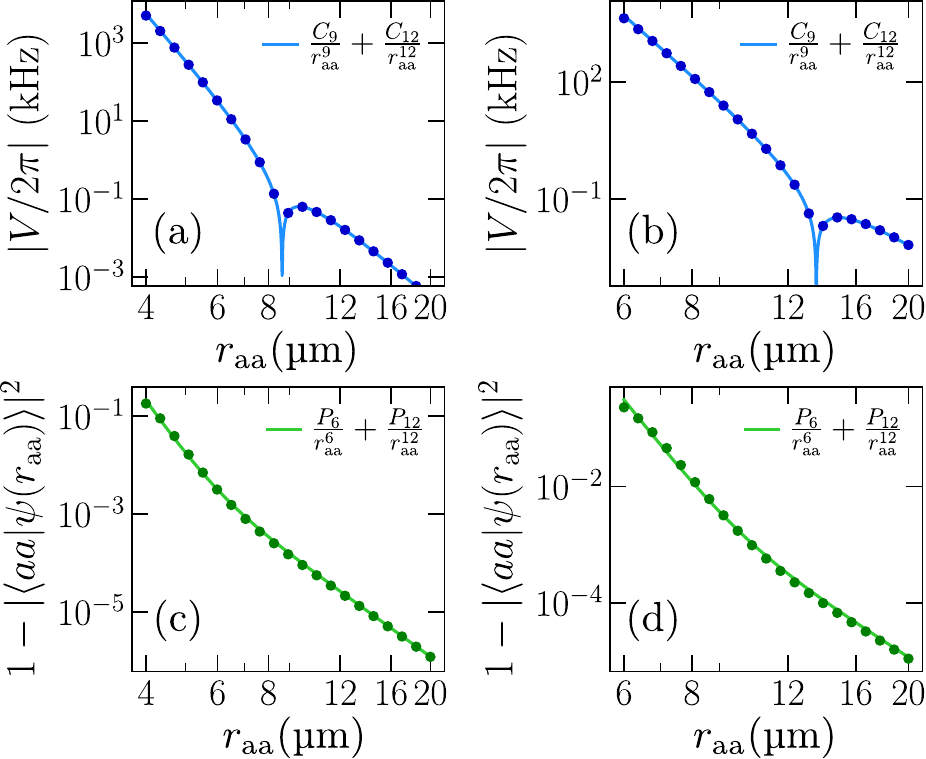} 
    \caption{Nullified vdW interactions for (a) NaCs and (b) RbCs \cite{Young2021,Weber2017}.  
    We plot the overlap of the corresponding two-atom eigenstate $|\psi(r_{\text{aa}})\rangle$ with $|aa\rangle$ for (c) NaCs and (d) RbCs. The lines denote fits using the perturbative values of $C_6 = 0, P_6$ (see End Matter) with $C_9, C_{12}, P_{12}$ as fitting parameters. For NaCs, $(C_9,C_{12})/2\pi = (\SI{163}{GHz.\micro\metre^9},\SI{-105}{THz.\micro\metre^{12}})$ and $(P_6,P_{12}) = ([\SI{2.06}{\micro\metre}]^6$$,[\SI{3.49}{\micro\metre}]^{12})$.    For RbCs, $(C_9,C_{12})/2\pi $$~=~$$ (\SI{4.8}{THz.\micro\metre^9},\SI{-12}{PHz.\micro\metre^{12}})$ and 
    $(P_6,P_{12}) = ([\SI{3}{\micro\metre}]^6,[\SI{5.42}{\micro\metre}]^{12})$.
    }
    \label{fig:Vr}
\end{figure}

In Fig.~\ref{fig:Vr}, we investigate the interactions for two examples where the Rydberg-Rydberg vdW interactions are nullified. We consider an example for each of NaCs and RbCs, the two species of bialkalis which have been trapped with tweezer arrays. In both cases, we detect with Cs atoms. The molecular states dressed are $|N=0,m_N = 0\rangle$ with $|N'=1, m_N'=1\rangle$ for NaCs and $|N=1, m_N=1\rangle$ with $|N'=2,m_N' = 2\rangle$ for RbCs. For both, the Rydberg states dressed are $|r\rangle = |n,L=1, J=3/2, m_J = 1/2\rangle, |r_\pi\rangle = |n, L=2, J=3/2, m_J = 1/2\rangle, |r_\sigma\rangle = |n-1, L=2, J=3/2, m_J = 3/2\rangle$, with principal quantum numbers $n=64$ for NaCs and $n=77$ for RbCs. In both cases, the atomic hyperfine structure is accounted for, and we ensure that the nullification is not due to a near-degeneracy of $|aa\rangle$ with another state causing one term in the perturbative expansion to dominate (see End Matter).

For NaCs, we use the Rydberg drive parameters $(\Omega_\pi, \Delta_\pi, \Omega_\sigma, \Delta_\sigma)/2 \pi \approx (-1250, 483, -348, 22.5)$ MHz and molecular drive $\Omega_{\text{m}}/2\pi =  -120$ kHz. For RbCs, we use the Rydberg drive parameters $(\Omega_\pi, \Delta_\pi, \Omega_\sigma, \Delta_\sigma)/2 \pi \approx (-668, 329, -196, 32.0)$ MHz and molecular drive $\Omega_{\text{m}}/2\pi = -12.9$ kHz. In both cases, $\Omega_\sigma$ is set to 10\% of the molecular transition frequency (and therefore approximately 10\% of the atomic transition frequency), while the $\Omega_\pi$ are both less than 4\% of the corresponding atomic transition frequency, ensuring that the rotating wave approximation holds. We use the Floquet approach of Ref.~\cite{Young2021} to diagonalize the interactions with a truncated Hilbert space. See End Matter for details on the truncated Hilbert space and the sensitivity of the nullified $C_3, C_6$ to the drive parameters.

The molecule-Rydberg interactions have strength $C_3 = 2\pi \times \SI{86}{kHz.\micro\metre^3} \ (2 \pi \times \SI{35}{kHz.\micro\metre^3})$ for NaCs (RbCs), and the Rydberg states $|a\rangle$ have lifetimes $\tau_{\text{a}} = \SI{225}{\micro\second}\ (\SI{391}{\micro\second})$ \cite{Robertson2021}. For atom-molecule separations of $\SI{1}{\micro\metre}$, the interaction time is 154 (110) times longer than the lifetime of the superposition $(|g\rangle + i e^{\pm i V_{\text{am}} t} |a\rangle)/\sqrt{2}$, so measurement errors from Rydberg decay are minimal. In both cases, both dipolar and vdW interactions have been nullified so that higher-order $C_9$ and $C_{12}$ coefficients become relevant at shorter distances. Likewise, we find the fraction of intermediate pair states involved, which describes the validity of the single-atom dressed-state basis and non-degenerate perturbation theory, has an additional $1/r_{\text{aa}}^{12}$ term. Nevertheless, the mixing remains perturbative at large distances until approximately \SI{4}{\micro\metre} (\SI{6}{\micro\metre}) for NaCs (RbCs), thereby setting the minimum length scale for the detection schemes.

At these minimal distances, the Rydberg-Rydberg interactions remain at the MHz scale but rapidly fall off as $r_{\text{aa}}$ increases since they are dominated by $C_{9}, C_{12}$ interactions, and at $r_{\text{aa}} = \SI{6}{\micro\metre} \ (\SI{9.5}{\micro\metre})$, they are on the scale of $2 \pi \times 10$ kHz for NaCs (RbCs), corresponding to reductions on the order of 50 compared to the typical dressed vdW interactions. In this regime, Rydberg blockade effects are easily overcome with sufficiently strong Rabi pulses. While the Rydberg-Rydberg interactions at these distances remain comparable to the molecule-Rydberg interactions at $\SI{1}{\micro\metre}$, the resulting interactions are sufficiently small that the techniques we consider below for various sources of crosstalk become suitable.

\emph{Remaining Crosstalk.}---When going beyond the case of a molecule-atom pair, many-body effects become important in the form of crosstalk. Crosstalk enters in three forms of decreasing strength: (i) Rydberg-Rydberg interactions, (ii) unwanted molecule-Rydberg interactions, and (iii) molecule-molecule interactions. In the previous sections, we discussed how (i) could be significantly mitigated, but not eliminated, through nullifying the dipolar and vdW interactions of the Rydberg state, with transient dipolar/vdW interactions depending on the precision of the drive parameters. In this section, we discuss general strategies for reducing the role of all three.

The most straightforward approaches are geometric manipulations and alternating measurements. In the former case, the distances involved in the unwanted interactions are increased, reducing their role, although this comes at the cost of decreasing any desirable interactions. Alternatively, the atom-molecule pairs could be spread apart for the measurement and then moved closer again afterward. In some instances, one can also utilize geometries where the strongest unwanted interactions fall along $1-3 \cos^2 \theta = 0$, further mitigating their role, although this does not work for vdW interactions. 

Likewise, by alternating measurements of the molecules with only some subset measured at a given step, the relevant length-scales of the unwanted interactions become larger at the cost of requiring multiple measurement cycles to measure all molecules. For example, in a linear chain with atoms positioned at the shortest perturbative distance, if we only perform the measurement on every other pair, the Rydberg-Rydberg distance is increased to $r_{\text{aa}} = \SI{8}{\micro\metre} \ (\SI{12}{\micro\metre})$, resulting in Rydberg-Rydberg interactions weaker than $2 \pi \times \SI{1}{kHz}$ for NaCs (RbCs), with another order of magnitude gained by increasing this to every three.

\begin{figure}
    \centering
    \includegraphics[width=1\linewidth]{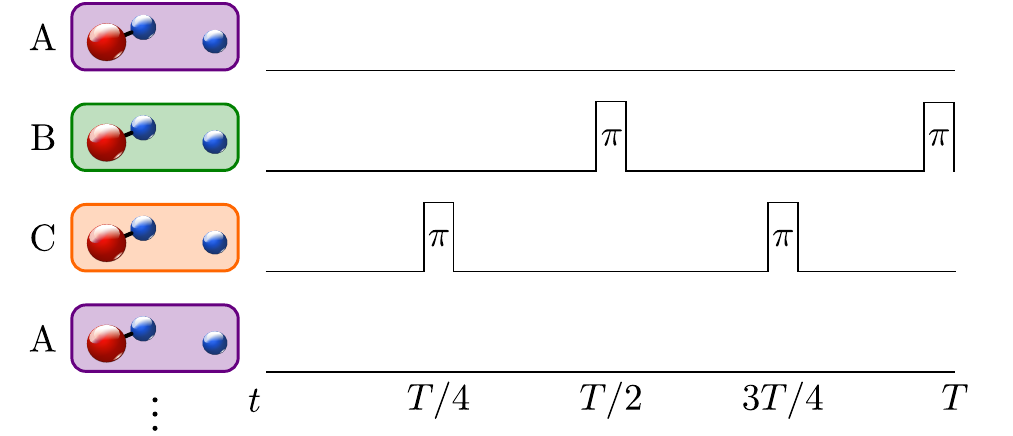}
    \caption{Spin-echo pulse sequence for reducing cross-talk. Only atom/molecules in same pulse class (A, B, C, ...) interact with one another, reducing unwanted crosstalk. In the above example, there are three classes.}
    \label{fig:pulse}
\end{figure}

Finally, a spin-echo approach may be utilized like in the state transfer protocol of Ref.~\cite{Guo2024}. This approach relies on the fact that the underlying interactions are Ising-like. In such a scenario, the effect of a $\pi$ pulse on a single particle corresponds to effectively flipping the sign of its Ising interactions with any other particle up to single-qubit rotations due to effective longitudinal field terms (i.e., single-particle $Z$ terms in the Hamiltonian). However, if two particles have a $\pi$ pulse simultaneously applied, then the sign does not change. For example, during the interaction time $T$, if one set of particles ``A'' undergoes no pulses and another set ``B'' has a $\pi$ pulse applied at time $T/2$, then the Ising-interactions between A and B are echoed out but the A-A and B-B interactions remain (an additional $\pi$ pulse is applied to B at time $T$ to return to the original basis). Additional pulse classes can be created through further subdivisions of $T$ by higher powers of two, see Fig.~\ref{fig:pulse}.

Through this approach, molecule-Rydberg pairs can be grouped into a total of $c > 1$ colors via a pulse sequence of $2^{c-1}$ total $\pi$ pulses, which for a 1D array of molecule-Rydberg pairs removes the unwanted interactions of the $c-1$ nearest pairs. Here, it is important that the Rydberg-Rydberg interactions are already reduced. While in principle even strong Rydberg-Rydberg interactions can be eliminated in this fashion, this relies on sufficiently strong Hadamard and $\pi$ pulses with higher susceptibility to errors if the pulses are not timed correctly.

\emph{Outlook.}---Our approach can readily be extended towards multi-molecule parity measurements. If each molecule interacts with $|a\rangle$ with strength $V$, an interaction time $t = \pi/V$ results in the accumulation of a parity-dependent $\pm 1$ factor in the relative phase of the superposition  $(|g\rangle + |a\rangle)/\sqrt{2}$, which can be measured via a similar pulse sequence ($H_Y \to H_X$). This would open possibilities for quantum error correction \cite{Devitt2013b,Terhal2015}, such as the surface code \cite{Bravyi1998, Dennis2002, Raussendorf2007, Fowler2012}, with a single entangling sequence for each measurement. Similarly, a spectroscopic approach can provide a means to simultaneously measure multiple molecules with a single Rydberg atom, although this is not scalable without additional atoms. Alternatively, the techniques here could also be extend towards gases of polar molecules in the absence of atom-molecule pairs, where the phase accumulated on the atom can nevertheless give insight into the local state(s) of the molecules. Finally, since our scheme only uses a single dressed Rydberg state, it provides a practical, less technically challenging first step that can then be extended to proposals based on multiple dressed Rydberg states \cite{Young2021,Bas}.

\begin{acknowledgments}
We thank A.~Safavi-Naini, B.~Gerritsen, and S.~Cornish for discussions. J.T.Y.~was supported by the NWO Talent Programme (project number VI.Veni.222.312), which is (partly) financed by the Dutch Research Council (NWO). A.V.G.~was supported in part by AFOSR MURI, ONR MURI, DoE ASCR Quantum Testbed Pathfinder program (awards No.~DE-SC0019040 and No.~DE-SC0024220), NSF QLCI (award No.~OMA-2120757), NSF STAQ program, ARL (W911NF-24-2-0107), DARPA SAVaNT ADVENT, and NQVL:QSTD:Pilot:FTL. A.V.G.~also acknowledges support from the U.S.~Department of Energy, Office of Science, National Quantum Information Science Research Centers, Quantum Systems Accelerator and from the U.S.~Department of Energy, Office of Science, Accelerated Research in Quantum Computing, Fundamental Algorithmic Research toward Quantum Utility (FAR-Qu). K.-K.N.~acknowledges support from AFOSR MURI (FA9550-20-1-0323).
\end{acknowledgments}

\bibliography{MoleculeRydberg,Extra}

\appendix

\onecolumngrid
\section{\large{End Matter}}
\vspace{-.3cm}
\twocolumngrid

\emph{Appendix A: Molecular hyperfine structure}.---In this appendix, we address the role of the hyperfine structure of the molecules. The hyperfine Hamiltonian is \cite{Aldegunde2017}
\begin{subequations}
\begin{equation}
    H_{\text{hf}} = H_Q + H_{IN} + H_{\text{t}} + H_{\text{sc}},
\end{equation}
\begin{equation}
\begin{aligned}
    H_Q = -e \sum_{i=1}^2 T^2(\nabla \mathbf{E}_i) \cdot T^2(\mathbf{Q}_i), \quad H_{IN} = \sum_{i = 1}^2 c_i \mathbf{N} \cdot \mathbf{I}_i, \\ 
    H_{\text{t}} = -c_3 \sqrt{6} T^2(\mathbf{C}) \cdot T^2(\mathbf{I}_1, \mathbf{I}_2), \quad H_{\text{sc}} = c_4 \mathbf{I}_1 \cdot \mathbf{I}_2,
\end{aligned}
\end{equation}
\end{subequations}
where $\mathbf{I}_i$ are the nuclear spins of the two alkali atoms, $e$ is the electric charge, $T^k_p$ is the $p$th component of a rank $k$ tensor, and $T^k_p(\mathbf{C}) \equiv C_p^k(\theta,\phi) = \sqrt{4 \pi/(2k+1)} Y_{k,p}(\theta,\pi)$ is a (renormalized) spherical harmonic. $H_Q$ is the quadrupole interaction, which can be written in the form $H_Q = \sum_{i=1}^2 (eQq)_i \frac{\sqrt{6}}{4 I_i (2 I_i -1)} T^2 (\mathbf{C}) \cdot T^2(\mathbf{I}_i, \mathbf{I}_i) $, where $eQ_i$ is the electric quadrupole moment and $q_i$ the negative of the electric field gradient of nucleus $i$. $H_{IN}$ is the spin-rotation coupling, and $H_{\text{t}}$ and $H_{\text{sc}}$ are the tensor and scalar interactions between the nuclear dipole moments. 

$H_Q, H_{IN}, H_\text{t}$ couple states with different $m_N$. In this case, strong microwave dressing of the molecule can render these processes off-resonant via state-dependent light shifts. However, given the smaller Rabi frequencies of the main text (limited by the Rydberg Rabi frequency), the inherent dressing is not always sufficient. In this case, additional microwave drives can be applied to the molecule. For example, for NaCs, a strong, resonant $\pi$ drive from $N=1$ to $N=2$ allows us to instead couple the $|N=0,m_N=0\rangle$ state to a superposition $(|N=1, m_N =1 \rangle + |N=2, m_N = 1\rangle)/\sqrt{2}$. Here, the additional microwave dressing is not limited by the Rydberg state, so strong light shifts can be applied. This comes with the caveat that the induced molecular dipole moment, and hence the molecule-Rydberg interaction, is reduced by a factor of $\sqrt{2}$. Likewise, if the two Rabi frequencies are comparable, it is important to ensure that the additional microwave dressing is adjusted such that the modified $|\pm\rangle$ states have no transition dipole moment between one another. 

Provided $m_N$ is conserved, then the remaining source of errors is due to differences in the nuclear spin eigenstates for the different rotational states. One possibility is to use strong Zeeman fields, although the magnetic field will modify the Rydberg states as well, which must be taken into account when mitigating the Rydberg-Rydberg interactions. Alternatively, if there are sets of hyperfine states whose eigenstates in the nuclear spin basis are largely independent of which rotational state they are in, these can likewise be used. For NaCs with the additional $\pi$ drive to the $N=2$ state, we find, in addition to the two nuclear-spin-stretched states, one additional set of states with an overlap of at least $\sim$ 97\% in the nuclear spin basis between the two pairs of coupled states. For RbCs, if the drive to reduce $m_N$ coupling is to the $|N=3, m_N=3\rangle$ state, we find, in addition to the nuclear-spin-stretched states, fourteen additional sets of states with an overlap of greater than 99\% in the nuclear spin basis between the two pairs of coupled states. This large number of states is, in part, a consequence of the significant nuclear scalar interaction in RbCs.

\emph{Appendix B: Dressing parameters}.---In this appendix, we present the analytic formulae for the dressing parameters in the case of the three-level Rydberg system. When the molecule is driven with ${\wp} = \pi$ polarization, the Rydberg dressing parameters for the $\sigma$ drive are
\begin{subequations}
    \begin{equation}
        \mp \Omega_\sigma = \frac{1 - \alpha^2 + 2 \mathcal{M}^2}{\Omega_\pi} - \frac{\sqrt{1 + 2 \mathcal{M}^2 }}{\sqrt{2 \mathcal{M}^2} \alpha}\Delta_\pi,
    \end{equation}
    \begin{multline}
        \Delta_\sigma = \frac{2 \mathcal{M}^2 \alpha^2-(1 + 2 \mathcal{M}^2)}{{2 \mathcal{M}^2 \alpha^2}} \Delta_\pi \\ +\frac{ (1 + 2 \mathcal{M}^2)^{3/2}(1- \alpha^2) }{2 \mathcal{M}^2 \alpha^3} \Omega_\pi,
    \end{multline}
\end{subequations}
while in the case where the molecule is driven with ${\wp} = \sigma$ polarization, the dressing parameters for the $\pi$ drive are
\begin{subequations}
    \begin{equation}
        \Omega_\pi = \pm \frac{1-\alpha^2 + (2 \mathcal{M}^2)^{-1}}{(\sqrt{2} \mathcal{M})^{-1} \alpha^2} \Omega_\sigma - \frac{ \sqrt{1+ (2 \mathcal{M}^2)^{-1} }}{\sqrt{2 \mathcal{M}^2}^{-1} \alpha} \Delta_\sigma,
    \end{equation}
    \begin{multline}
        \Delta_\pi = \frac{(2 \mathcal{M}^2)^{-1} \alpha^2 - (1 + (2 \mathcal{M}^2)^{-1})}{2 \mathcal{M} \alpha^3} \Delta_\sigma \\ \pm \frac{(1 + (2 \mathcal{M}^2)^{-1})^{3/2}(1-\alpha^2)}{(2 \mathcal{M}^2)^{-1} \alpha^3} \Omega_\sigma.
    \end{multline}
\end{subequations}
We note that the equations are invariant under the transformation $\alpha \to - \alpha, \Omega_\sigma \to -\Omega_\sigma$, corresponding to $|r_\sigma\rangle \to - |r_\sigma\rangle$, so we consider only $\alpha > 0$. Furthermore, the two possible solutions can be captured via a single solution by simply tuning the sign of $\Omega_\sigma$, so without loss of generality we use the $(\pm, \mp) \to (+,-)$ solution in the main text.

\emph{Appendix C: Rydberg fine structure}.---When the drive is comparable or stronger than the Rydberg fine structure, it will couple to additional states, modifying the interactions and therefore the dressing parameters needed to remove the Rydberg interactions. We neglect the hyperfine structure, which is several orders of magnitude smaller than the Rabi frequencies. For each of the original states $|r\rangle, |r_\pi\rangle, |r_\sigma\rangle$, an additional state which differs only in $J$ will be coupled (due to the electron spin $S=1/2$), which we denote $|r'\rangle, |r_\pi'\rangle, |r_\sigma'\rangle$, with the exception that there will be no additional primed state if the corresponding unprimed state is an S$_{1/2}$ state. We thus write $|a\rangle \propto |r\rangle + a|r'\rangle + a_\pi |r_\pi\rangle + a_\pi' |r_\pi \rangle + a_\sigma |r_\sigma\rangle + a_\sigma' |r_\sigma' \rangle$.

The resulting microwave-dressing Hamiltonian is
\begin{subequations}
\begin{equation}
    H_{\text{mw}}' = \left( 
    \begin{array}{cccccc}
        0 & 0 & \Omega_\pi & \Omega_\sigma & f_\pi \Omega_\pi & f_\sigma \Omega_\sigma \\
        0 & -\Delta & g_\pi \Omega_\pi & g_\sigma \Omega_\sigma & g_\pi' \Omega_\pi' & g_\sigma' \Omega_\sigma \\
        \Omega_\pi & g_\pi \Omega_\pi & - \Delta_\pi & 0 & 0 & 0\\
        \Omega_\sigma &  g_\sigma \Omega_\sigma &  0 & - \Delta_\sigma & 0 & 0\\
        f_\pi \Omega_\pi & g_\pi' \Omega_\pi & 0 & 0 & -\Delta_\pi' & 0 \\
        f_\sigma  \Omega_\sigma & g_\sigma' \Omega_\sigma & 0 & 0 & 0 & -\Delta_\sigma'
    \end{array}
    \right),
\end{equation}
\begin{equation}
    f_\pi = \frac{\langle r_\pi'| d_0| r \rangle}{\langle r_\pi| d_0| r \rangle},
    \qquad
    f_\sigma = \frac{\langle r_\sigma'| d_+| r \rangle}{\langle r_\sigma| d_+| r \rangle},
\end{equation}
\begin{equation}
    g_\pi = \frac{\langle r_\pi| d_0| r' \rangle}{\langle r_\pi| d_0| r \rangle},
    \qquad
    g_\sigma = \frac{\langle r_\sigma| d_+| r' \rangle}{\langle r_\sigma| d_+| r \rangle},
\end{equation}
\begin{equation}
    g_\pi' = \frac{\langle r_\pi'| d_0| r' \rangle}{\langle r_\pi| d_0| r \rangle},
    \qquad
    g_\sigma' = \frac{\langle r_\sigma'| d_+| r' \rangle}{\langle r_\sigma| d_+| r \rangle},
\end{equation}
\begin{equation}
    \Delta' = E_{r'}-E_{r},
    \quad
    \Delta_\pi = E_{r_\pi'} - E_{r},
    \quad
    \Delta_\sigma = E_{r_\sigma'} - E_{r}.
\end{equation}
\end{subequations}
In this case, there are no dipolar Rydberg interactions if
\begin{multline}
     |a_\pi + f_\pi a_\pi' + a (g_\pi a_\pi + g_\pi' a_\pi')|^2 \mu_\pi^2 - \\ |a_\sigma + f_\sigma a_\sigma' + a (g_\sigma a_\sigma + g_\sigma' a_\sigma')|^2 \mu_\sigma^2/2 = 0.
\end{multline}
Although additional states have been introduced, leading to further constraints (to ensure $|a\rangle$ is an eigenstate), corresponding additional degrees of freedom (in the state coefficients) are simultaneously introduced. Unlike for three levels, the drive parameters cannot be determined analytically, so numerical solutions must be found. We achieve this in an adiabatic manner, ramping the molecular drive strength while using the free parameters for the previous step as initial guesses for the next. Hence in the limit of weak drive, this reduces to the three-level solution. For this numerical approach, we only allow the detuning of the drive which does not couple the molecular states to be varied compared to the three-level value, while the other dressing parameters are kept the same. 

\emph{Appendix D: vdW nullification}.---In this section, we illustrate the identification of parameters which nullify the vdW interactions. We use the perturbative Floquet approach of Ref.~\cite{Young2021} to determine $C_6$ at $\theta = \pi/2$ using the atomic data from Ref.~\cite{Weber2017}. Furthermore, we restrict the parameters we investigate by fixing the molecular Rabi frequency to correspond to a Rydberg Rabi frequency which is 10\% of the molecular transition frequency. This is done to achieve a large molecular Rabi frequency so that the role of the hyperfine structure can be reduced. 

\begin{figure}[t]
    \centering
    \includegraphics[width=0.49\linewidth]{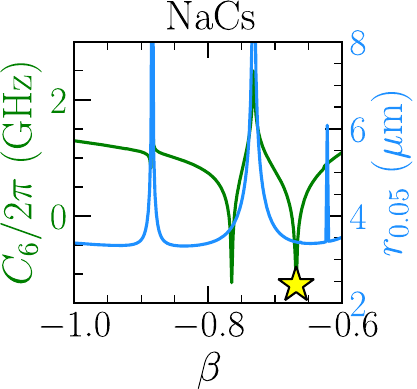}
    \includegraphics[width=0.49\linewidth]{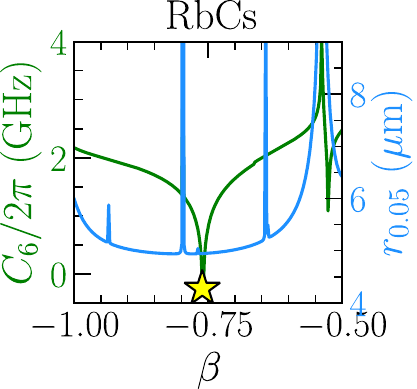}
    \caption{Perturbative vdW nullification for NaCs (left) and RbCs (right). Stars denote the parameters used in the main text. For NaCs (RbCs), the Hilbert space is truncated to $L\leq 3$, $60 (73) \leq n \leq 67 (80)$, and pair-state energies up to $2 \pi \times \SI{16}{GHz}$ ($2 \pi \times \SI{9}{GHz}$) away from the dressed-state energies.}
    \label{fig:endC6}
\end{figure}

Once the molecular Rabi frequency has been fixed, $\beta$ is the only free parameter. In Fig.~\ref{fig:endC6}, we show $C_6$ as a function of $\beta$. We also quantify the extent to which the dipolar interactions remain perturbative via $r_{0.05}$, which is the distance at which the pair-state overlap $1-|\langle a a|\psi(r_{0.05})\rangle|^2 \approx P_6/r_{0.05}^6 = 0.05$ according to the perturbative coefficient $P_6$ (higher-order terms are neglected). This ensures that the vdW nullification is not due to a resonance making a single term strong, so pairs of atoms, and thus atom-molecule pairs, can remain close.

An important consideration is the sensitivity of the nullification odipolar and vdW interactions to imperfections in the dressing parameters. We quantify the fractional sensitivity for $\mathbf{v} \equiv (\Omega_\pi, \Omega_\sigma, \Delta_\pi, \Delta_\sigma)$ via $\mathbf{J}_p \equiv \mathbf{v} \odot \partial_\mathbf{v} C_\alpha$, where $\alpha = 3,6$ denote dipolar and vdW coefficients, respectively, and $\odot$ denotes element-wise multiplication. For NaCs, we find $\mathbf{J}_3 = 2 \pi \times (123,-114,43.9,-1.57)~\SI{}{\mega\Hz.\micro\metre^3}$ and $\mathbf{J}_6 = 2 \pi \times (175,12.9,5.47,-0.118)~\SI{}{\giga\Hz.\micro\metre^6}$, with the greatest uncertainty from $\Omega_\pi$. For 0.1\% fractional error in $\Omega_\pi$ at $\SI{4}{\micro\metre}$ ($\SI{8}{\micro\metre}$), the excess interaction is $2\pi \times \SI{45}{kHz}$ ($2\pi \times \SI{1}{kHz}$). For RbCs, we find $\mathbf{J}_3 = 2 \pi \times (232,-237,115,-8.45)~\SI{}{\mega\Hz.\micro\metre^3}$ and $\mathbf{J}_6 = 2 \pi \times (356,24.3,29.9,-0.035)~\SI{}{\giga\Hz.\micro\metre^6}$, with the greatest uncertainty from $\Omega_\pi$ once more. For 0.1\% fractional error in $\Omega_\pi$ at $\SI{6}{\micro\metre}$ ($\SI{12}{\micro\metre}$), the excess interaction is $2\pi \times \SI{8.7}{kHz}$ ($2\pi \times \SI{0.25}{kHz}$).

Finally, we determine the effect of $\phi$ and complex Rabi frequencies on the values of $C_6, C_9, C_{12}$. Because the dressed states are superpositions of multiple states, virtual processes between these states in a perturbative expansion can introduce phase dependencies (e.g., $|rr\rangle \to |r_\sigma r_\sigma\rangle$) for the dressed state interaction, with the drive phases and $\phi$ playing a similar role. For the above truncated Hilbert spaces, we find no variation in $C_6$ according to the perturbative calculation. If the truncated Hilbert space Hamiltonians are diagonalized, we find that $C_9$ varies by no more than 2\% and $C_{12}$ has negligible variations across several choices of $\phi$.

\end{document}